\documentclass[sigconf,natbib=false,authorversion,nonacm]{acmart}

\usepackage{amsmath,amsfonts}
\usepackage{algorithmic}
\usepackage{algorithm}
\usepackage{array}
\usepackage[caption=false,font=normalsize,labelfont=sf,textfont=sf]{subfig}
\usepackage{textcomp}
\usepackage{stfloats}
\usepackage{url}
\usepackage{verbatim}
\usepackage{graphicx}
\usepackage{bm}
\usepackage{xcolor}
\usepackage{numprint}
\usepackage{datatool}
\usepackage{hyperref} 
\usepackage[normalem]{ulem} 
\usepackage{siunitx}
\usepackage{tikz}
\usetikzlibrary{arrows,arrows.meta,bending,calc}
\usetikzlibrary {positioning}
\usetikzlibrary{shapes,snakes}
\usepackage{listings}
\usepackage[endianness=big]{bytefield}
\usepackage{booktabs}
\usepackage{arydshln} 
\usepackage[shortcuts]{extdash}

\AtBeginDocument{%
  }

\setcopyright{acmlicensed}

\copyrightyear{2026}
\acmYear{2026}
\acmDOI{XXXXXXX.XXXXXXX}
\acmConference[SAC'26]{The 41st ACM/SIGAPP Symposium on Applied Computing}{March 23--27, 2026}{Thessaloniki, Greece}
\acmISBN{979-X-XXXX-XXXX-X/26/03}

\RequirePackage[
  datamodel=acmdatamodel,
  style=acmnumeric,
  ]{biblatex}

\newcommand{\bitssubclass}{\color{lightgray}\rule{\width}{\height}}

\newcommand{\emad}[1]{{\color{red} Emad: #1}}

\renewcommand{\emad}[1]{}

\definecolor{bblue}{HTML}{4F81BD}
\definecolor{rred}{HTML}{C0504D}
\definecolor{ggreen}{HTML}{9BBB59}
\definecolor{ppurple}{HTML}{9F4C7C}

\newcommand{\sixteenBitSpace}{65536}
\newcommand{\thirtyTwoBitSpace}{4294967296}
\newcommand{\scryOccupancy}{\fpeval{round((\ScryPoints/\sixteenBitSpace)*100)}}
\newcommand{\riscvOccupancy}{\fpeval{round((\riscvPointsRelevant/\thirtyTwoBitSpace)*100)}}

\lstdefinelanguage{scryasm}{
	keywords={},
	otherkeywords={Enlarge, High, Low, u8, i8, u16, i16, u32, i32, u64, i64, Private},
	keywords = [2]{const, grow, add, add.s, sub, sub.s, pick, pick.i, echo, echo.l, dup, jmp, mul, div, shr, shl, ld, ld.s, st, st.s, call, ret, rsrv, free, lt, gt, eq, nop, and, or, xor},
	keywords = [3]{..},
	alsoletter={.}
	sensitive=false, 
	morecomment=[l]{//}, 
	morecomment=[s]{/*}{*/}, 
	morestring=[b]" 
}
\lstdefinestyle{scryasmstyle}{
	float=t,
	belowskip=-0.3cm,
	frame=single, 
	frameround=tttt,
	language=scryasm,
	basicstyle=\footnotesize\ttfamily,
	commentstyle=\itshape\color{green!60!black},
	keywordstyle=[2]\color{blue!80!black}\bfseries,
	keywordstyle=[3]\color{black}\bfseries,
	keywordstyle=\color{red!80!black},
	tabsize=4,
	numbers=left,
	numbersep=8pt,
	stepnumber=1,
	numberstyle=\tiny\color{gray},
	captionpos=b,	
	linewidth=.47\textwidth,
	xleftmargin=0.4cm,
	escapechar = !,
	identifierstyle=\color{orange}\bfseries,
}
\newcommand{\gnode}[1]{\tikz[remember picture, overlay] \node [] (#1){};}

\usepackage{draftwatermark}
\SetWatermarkText{DRAFT}
\SetWatermarkScale{1}
\SetWatermarkLightness{0.9}

\begin{document}

\title{A Dense and Efficient Instruction Set Architecture Encoding}

\author{Emad Jacob Maroun}
\email{ejama@dtu.dk}
\orcid{0000-0002-3675-3376}
\affiliation{%
  \institution{Technical University of Denmark, Department of Applied Mathematics and Computer Science}
  \city{Kongens Lyngby}
  \country{Denmark}
}

\renewcommand{\shortauthors}{E. J. Maroun}

\newcommand{\codePointslui}{33554432}
\newcommand{\codePointsauipc}{33554432}
\newcommand{\codePointsjal}{33554432}
\newcommand{\codePointsjalr}{4194304}
\newcommand{\codePointsbeq}{4194304}
\newcommand{\codePointsbne}{4194304}
\newcommand{\codePointsblt}{4194304}
\newcommand{\codePointsbge}{4194304}
\newcommand{\codePointsbltu}{4194304}
\newcommand{\codePointsbgeu}{4194304}
\newcommand{\codePointslb}{4194304}
\newcommand{\codePointslh}{4194304}
\newcommand{\codePointslw}{4194304}
\newcommand{\codePointslbu}{4194304}
\newcommand{\codePointslhu}{4194304}
\newcommand{\codePointssb}{4194304}
\newcommand{\codePointssh}{4194304}
\newcommand{\codePointssw}{4194304}
\newcommand{\codePointsaddi}{4194304}
\newcommand{\codePointsslti}{4194304}
\newcommand{\codePointssltiu}{4194304}
\newcommand{\codePointsxori}{4194304}
\newcommand{\codePointsori}{4194304}
\newcommand{\codePointsandi}{4194304}
\newcommand{\codePointsadd}{32768}
\newcommand{\codePointssub}{32768}
\newcommand{\codePointssll}{32768}
\newcommand{\codePointsslt}{32768}
\newcommand{\codePointssltu}{32768}
\newcommand{\codePointsxor}{32768}
\newcommand{\codePointssrl}{32768}
\newcommand{\codePointssra}{32768}
\newcommand{\codePointsor}{32768}
\newcommand{\codePointsand}{32768}
\newcommand{\codePointsfence}{4194304}
\newcommand{\codePointsecall}{1}
\newcommand{\codePointsebreak}{1}
\newcommand{\codePointsmul}{32768}
\newcommand{\codePointsmulh}{32768}
\newcommand{\codePointsmulhsu}{32768}
\newcommand{\codePointsmulhu}{32768}
\newcommand{\codePointsdiv}{32768}
\newcommand{\codePointsdivu}{32768}
\newcommand{\codePointsrem}{32768}
\newcommand{\codePointsremu}{32768}
\newcommand{\codePointslwu}{4194304}
\newcommand{\codePointsld}{4194304}
\newcommand{\codePointssd}{4194304}
\newcommand{\codePointsslli}{65536}
\newcommand{\codePointssrli}{65536}
\newcommand{\codePointssrai}{65536}
\newcommand{\codePointsaddiw}{4194304}
\newcommand{\codePointsslliw}{32768}
\newcommand{\codePointssrliw}{32768}
\newcommand{\codePointssraiw}{32768}
\newcommand{\codePointsaddw}{32768}
\newcommand{\codePointssubw}{32768}
\newcommand{\codePointssllw}{32768}
\newcommand{\codePointssrlw}{32768}
\newcommand{\codePointssraw}{32768}
\newcommand{\codePointsmulw}{32768}
\newcommand{\codePointsdivw}{32768}
\newcommand{\codePointsdivuw}{32768}
\newcommand{\codePointsremw}{32768}
\newcommand{\codePointsremuw}{32768}

\newcommand{\riscvOpcodesAll}{878}
\newcommand{\riscvOpcodesRelevant}{121}
\newcommand{\riscvPointsAll}{4285375309}
\newcommand{\riscvPointsRelevant}{2902171650}
\newcommand{\ScryPoints}{18072}

\begin{abstract}
	Instruction density and encoding efficiency are some of the few things directly affected by an instruction set architecture's design.
	In contrast, a processor's implementation often significantly influences performance, power efficiency, and area usage.
	Therefore, a major goal of instruction set design should be maximizing instruction density and encoding efficiency.
	This paper introduces the design elements of the Scry instruction set architecture that most significantly affect instruction density and encoding efficiency.
	Scry is a novel and experimental instruction set that revisits first principles to design an instruction set fit for modern processor implementations.
	
	Scry uses forward-temporal referencing as a means of data flow, where instructions refer to which future instructions consume their outputs.
	It also uses internal tagging, where the processors track data types internally, to reduce the number of instructions needed and increase flexibility.
	Combining these two methods, Scry achieves instruction-feature parity with RISC\=/V's RV64IMC using only 2-byte instructions compared to RISC\=/V's 4 bytes.
	Scry's instructions occupy only \scryOccupancy\% of the 2-byte encoding space, where RV64IMC instructions occupy \riscvOccupancy\% of the 4-byte encoding space.
	We show that hand-compiled Scry's static instruction density is comparable to RV64IMC for small functions and improves as functions grow in size.
	
\end{abstract}



\maketitle

\section{Introduction}

An instruction set architecture (ISA) defines the basic instructions processors understand and execute.
Some instructions accept inputs to perform some operations and produce outputs used in subsequent instructions.
This is called \textit{data flow}.
Instructions are typically executed in order.
Some instructions change which instruction should be executed next, with the sequence continuing from the new instruction.
This is called \textit{control flow}.
The design of an ISA comprises data and control flow instructions, combined with other miscellaneous instructions, such that computations are performant and efficient.

Almost all modern ISAs follow a similar design.
They use registers for data flow, where instructions refer to which registers hold their input data and which register(s) should hold their outputs.
Modern processor implementations use sophisticated techniques to maximize performance while maintaining the semantics of the program.
They try to execute multiple instructions in parallel and reorder them so instructions with high latencies do not cause bottlenecks~\cite{hennessy06computer}.
These efforts are hindered by the decades-old design of contemporary ISAs.
The limited number of available registers in the ISA (architectural registers) means \textit{false dependencies} arise when more operands are live than there are registers in the ISA.
Register renaming circumvents the ISA limit on registers if more registers are available in a given processor implementation~\cite{sima2000design}.
Renaming allows operands to be stored in different physical registers even when using the same architectural register.
This allows the processor to increase parallel execution.
However, it is a complicated process, which consumes a significant portion of a processor's power~\cite{kucuk2003energy}, limits a processor's frequency~\cite{palacharla1997complexity}, and limits the number of instructions that can be simultaneously checked for parallelism---reducing its exploitation~\cite{petit2013efficient}.
Methods for data flow that do not depend on registers can also avoid the problems of false dependencies and register renaming.

To perform operations on different data types, like signed or unsigned integers or floating-point numbers (floats), distinct instructions are provided to handle each data type.
However, as ISAs mature, they must be extended with additional functionality and, therefore, additional instructions.
Every extension that adds a data type must include additional instructions that perform existing operations on that new type.
Mature ISAs, therefore, have a problem with running out of space for new instructions.
For example, the x86 ISA has steadily increased the number of instructions (to now well over a thousand) and their average size (the newest averaging close to four bytes per instruction)~\cite{lopes2015shrink}.
The relatively young RISC\=/V ISA would use 99.78\% of its available encoding space if it did not reuse some code points between the 32- and 64-bit ISAs~\cite{maroun2025internally}.
This leaves little room for future extensions and limits how much other organizations can customize the ISA for their use cases---a core selling point of RISC\=/V.

\textit{Scry} is a new ISA that aims to support modern processor implementations in their quest for performance.
This first paper on the ISA describes its design with a focus on its encoding.
While it is often difficult to establish the degree to which the ISA impacts often-used performance metrics~\cite{akram2017impact}, metrics that can be attributed to the ISA are instruction density (how many bytes of instruction data are needed to perform a task) and encoding efficiency (how much of the encoding space is used). 
Therefore, ensuring that an ISA design is conducive to dense code and an efficient encoding is important~\cite{weaver2009code}.
It also aims not to use registers as a means of data flow, to avoid implementations using register renaming and thereby incurring its costs.
This paper will describe the major features of the ISA and how they affect the encoding.
While short arguments for how some features benefit the performance and implementation of processors will be given, these topics are otherwise out of the scope of this paper.

Scry has two major features differentiating it from traditional ISAs that this paper covers:
Data flow uses \textit{forward-temporal referencing} to pass operands from producers to consumers without explicit registers.
Forward-temporal referencing does not exhibit false dependencies and does not necessitate register renaming in high-performance processors.
This referencing scheme only requires instructions to specify where its outputs are used.
Since most instructions have fewer outputs than inputs, this minimizes the encoding space needed for data flow.
Forward-temporal referencing also enables varying semantics based on the number of instruction inputs, further increasing encoding efficiency.

\textit{Internal tagging} tracks data types in the processor, removing the need for type-specific instructions and enabling varying instruction semantics based on type.
This significantly reduces the necessary number of instructions and allows for additional semantics, based on instruction input types, at no additional encoding cost.
The effect of these features is an ISA that is orders of magnitude more encoding efficient than traditional ISAs.
It has only 16-bit instructions and a feature-set equivalent to RISC\=/V's RV64IMC, with ample opportunity for extension.

This paper is organized into seven sections:
The following section covers related work within ISA design and encoding density.
\autoref{sec-forward} describes how forward-temporal references manage data flow and enable operand-count-polymorphism.
\autoref{sec-tagged} describes how Scry uses internal tagging and type-polymorphism.
\autoref{sec-encoding} presents the encoding of all the Scry instructions into 16-bit words.
\autoref{sec-evaluation} evaluates Scry's encoding efficiency and static instruction density compared to RISC\=/V.
\autoref{sec-conclusion} concludes. 

\section{Related Work}

Significant work has been done in alternative execution and data flow methods to increase performance and efficiency.

Dataflow computing (DFC) architectures are different from traditional control-flow computing (CFC) architectures in that instruction order is irrelevant~\cite{lee1993issues, hurson2007dataflow}.
Instead, outputs are directly assigned to the consuming instructions (\textit{forward referencing}) using their addresses as identifiers (\textit{spatial referencing}).
Forward referencing makes parallelism explicit and easy to identify and exploit: all instructions with ready inputs may execute in parallel. 
Inherent inefficiencies in DFC mean it has not become mainstream~\cite{yazdanpanah2013hybrid}.
Spatial referencing makes the instruction stream unpredictable and requires repeatedly comparing operands to identify ready instructions, resulting in high overhead~\cite{watson1982practical,petersen2006reducing}.
Pure DFC architectures also have inherent disadvantages when targeting imperative languages (like C)~\cite{yazdanpanah2013hybrid}, while modern superscalar CFC approaches are superior in specific cases~\cite{budiu2005dataflow}.
Hybrid and heterogeneous architectures try to balance DFC and CFC by supporting them in one processor~\cite{yazdanpanah2013hybrid, nowatzki2019heterogeneous}.
Some have attempted to divide program execution to use the best paradigm for the task~\cite{kavi2001scheduled,burger2004scaling,sharma2010graceful} or adding DFC features to established CFC architectures~\cite{cowley2021risc}.
Scry leverages the parallelism of DFC's forward referencing but avoids the pitfalls of spatial referencing.
Instructions are executed in traditional, sequential order, while \textit{temporal referencing} exploits parallelism without the need for complicated operand management.

The STRAIGHT ISA is also designed to eliminate false dependencies and the need for register renaming~\cite{irie2018straight}.
It takes a similar approach to Scry using \textit{backwards-temporal} references.
Each instruction specifies when its operands were produced without specifying when its outputs are used.
A register file stores operands in a queue of the length of the maximum reference distance.
When an instruction is executed, the queue is advanced once, discarding the front of the queue as its value can no longer be referenced.
STRAIGHT had strong constraints on instruction placements, e.g., around branches and loops, resulting in large increases in instruction counts.
The Clockhands ISA alleviates the issue by introducing additional operand queues that only advance when a value is pushed to them individually~\cite{koizumi2023clockhands}. 
This allows the compiler to handle values with various lifetime characteristics more efficiently. 
The implementation uses four queues (what is called ``Hands''), one for temporaries, one for long-lived values, one for loop constants, and one for stack pointers and function arguments.
The additional queues allow Clockhands to reduce instruction counts, putting it on par with RISC\=/V instruction counts.
Scry's use of forwards- instead of backwards-temporal references gives the processor information about operand lifetimes.
It also makes instruction dependency resolution easier, needing only to check if an instruction outputs directly to a succeeding instruction to determine if they can be executed in parallel.

Tagged architectures have been extensively studied.
Traditionally, tagged architectures store tags adjacent to their data in both the processor and memory~\cite{feustel1973advantages}.
Their benefits included simplifying software, supporting programming languages, compilers, operating systems (OSs), and achieving better register utilization, data scheduling, and parallelism of functional units~\cite{feustel1971rice}.
However, their benefits likely did not outweigh their overheads~\cite{gehringer1985tagged}.
For example, tags supporting custom types could double the memory footprint~\cite{hagelberg2005tagged}.
Modern tagged architectures have focused on various types of safety~\cite{jero2022tag}. 
Internally tagged architectures were proposed to provide some of the traditional benefits without the overheads~\cite{maroun2025internally}.

\section{Forward-Temporal Referencing}
\label{sec-forward}

In most ISAs, data flow is handled using registers: 
Instructions fetch their inputs from predetermined registers and output their result to the same or other predetermined registers.
Initial instructions output their results to registers named in the instruction's encoding.
The current instruction reads its inputs from named registers and outputs its result to a named register.
Following instructions, name the register to consume the current instruction's result.

To maximize performance, processors try to execute several instructions in parallel.
Register-based referencing complicates and limits parallel execution.
Because registers are limited (often only 32 or 64 are available), parallelism is lost when two independent instructions must write to the same register.
This \textit{false dependency} limits the processor's ability to execute instructions simultaneously.
Contemporary processors must detect and circumvent false dependencies to achieve high performance.
Register renaming eliminates false dependencies by using additional registers~\cite{sima2000design}.
Each instruction's logical operands are mapped (renamed) to physical registers before execution.
False dependencies are broken by assigning independent operands to different physical registers when they use the same logical registers.
Renaming is a complicated process, which consumes a significant portion of a processor's power~\cite{kucuk2003energy}, limits a processor's frequency~\cite{palacharla1997complexity}, and limits the number of instructions that can be simultaneously checked for parallelism---reducing its exploitation~\cite{petit2013efficient}.

To avoid false dependencies and register renaming, Scry uses a novel data flow scheme we call \textit{forward-temporal referencing.}
Instead of using named resources for data flow---which are necessarily limited---Scry uses references to other instructions as data flow.
Initial instructions specify when their outputs are consumed.
When execution reaches an instruction, the inputs it needs to operate on are already specified by preceding instructions.
Scry's references are temporal, describing when the operands will be consumed and not the position of the consuming instruction.
For example, a reference value of 0 means the next instruction in the instruction stream will consume the operand, a 1-reference means the second instruction will consume it, etc.
If execution branches at runtime, different instructions will be executed, and the operands will therefore automatically flow to the executed instruction.
Each instruction only refers to when its output(s) will be consumed, with its inputs being implicit.
The processor is responsible for managing the lifetime of operands so that they arrive at the functional units when needed.
\begin{lstlisting}[style=scryasmstyle, caption={Example Scry assembly with reference target highlight.}, label=lst-flow-example]
	!\gnode{lst-flow-example-e}!
	!\gnode{lst-flow-example-a}!add.s =>3    // First operand 
	nop
	!\gnode{lst-flow-example-b}!add.s =>1    // Second operand
	nop
	!\gnode{lst-flow-example-c}!add.s =>10   // add first and second
	!\gnode{lst-flow-example-d}!
\end{lstlisting}
\begin{tikzpicture}[remember picture, overlay, green!60!black]
	\path[->, bend right] 
	(lst-flow-example-e.north east) edge (lst-flow-example-a) 
	(lst-flow-example-e.north west) edge (lst-flow-example-b) 
	(lst-flow-example-a) edge (lst-flow-example-c)
	(lst-flow-example-b) edge (lst-flow-example-c)
	(lst-flow-example-c) edge (lst-flow-example-d.south)
	;
\end{tikzpicture} 

\autoref{lst-flow-example} shows an example Scry assembly program using three addition instructions with highlighted reference targets (green arrows) matching the output argument.
The first and second additions take implicit inputs from preceding instructions.
They reference the third addition with an offset argument(\texttt{=>3} and \texttt{=>1}) equal to the number of instructions between the producer and consumer in the instruction stream: 3 and 1, respectively.
The ``temporal'' in Scry's references refers only to the number of instructions executed and not clock cycles or the specific position of instructions in the binary.
I.e. \texttt{=>3} means whatever instruction is the fourth to be executed after the current one, will be the consumer of the output.
In \autoref{lst-flow-example}, had there been a branch instruction before the third addition, whatever instruction was first in the new instruction stream would have consumed the first two additions' outputs.
The order of references is meaningful.
Each instruction's inputs are ordered, with the operands produced by earlier instructions being first in the order.
This could be important for instructions like \texttt{sub}, where the second operand (i.e., produced later) is subtracted from the first.
It is the programmer's/compiler's responsibility to order instructions such that their outputs reach the consuming instruction in the correct order.
The dedicated data flow instructions can be leveraged to ensure a correct order can always be found.

In the Scry assembly language, we distinguish between an instruction's \textit{arguments}, which are statically encoded into the instruction (e.g., an output reference ``\texttt{=>3}''), and \textit{operands}, which are the runtime inputs of an instruction and are not statically determined by the encoding.
The third addition in \autoref{lst-flow-example} has one argument (\texttt{=>10}) and two operands from the previous additions.
The two initial additions also have one argument each (\texttt{=>3} and \texttt{=>1}) but only one operand, as can be seen from their green incoming arrows.
This also illustrates that the number of operands that reach an instruction, and their order, is decided by the output references of preceding instructions and can therefore vary between programs.

Because output references are given statically as arguments, they cannot be used where the consumer is a variable distance from the producer.
For example, there could be a loop between the producer and consumer instructions, which could iterate any number of times.
A reference output cannot account for such variability.
Instead, the program stack must be used to store the operand and reload it when needed.
Scry has first-class support for accessing and managing the stack, but they are out of the scope of this paper.

\subsection{Data Flow Instructions}

Fine-grained control of operands is necessary to ensure Scry can support existing coding patterns.
Scry, therefore, has dedicated data flow instructions whose sole purpose is managing operands so that they reach their desired consumer.

Most Scry instructions only have one output reference.
This means only one instruction can consume the value.
However, some values must be used multiple times. 
For example, loop counters are needed once to check if the iteration must continue and once to increment or decrement the counter for the next iteration.
Therefore, Scry provides an instruction for duplicating operands.
This is preferable to giving all instructions additional output references.
The authors of~\cite{franklin1992register} show that a vast majority of values are used only once, with only very few being used more than four times.
The authors of~\cite{tseng2008achieving} present similar statistics on the SPEC CPU2000 benchmark suite.
There, 70\% of values are only used once, while over 90\% are used only twice.

Scry's default output references reach only 32 instructions ahead, saving on encoding space by only needing 5 bits.
This is a relatively short reach.
However, the authors of~\cite{van2010managing} show that 90\% of values live for less than 11 clock cycles.
The authors of~\cite{butts2004use} and~\cite{hu2000reducing} both show that the active lifetime of values (the time from being written to a register to the last read) is in the low single digits.
Lastly,~\cite{lozano1995exploiting} shows that up to 95\% of all values will be dead within the 32 instructions following their write. \emad{This paper also shows that short-lived values mostly do not cross basic block boundaries, meaning most values do not cross the boundary.}
Therefore, for the vast majority of operands, a reach of 32 should be enough.
However, we provide a dedicated instruction for the few operands that must live longer.
Providing dedicated duplication and long-reach instructions is the most efficient way of supporting multiple uses and long lifetimes of operands at the lowest encoding space usage.

\textbf{Echo:} Scry's first primary data flow instruction retargets in-flight operands to where they need to go.
The first variant is the \textit{long echo}, whose sole purpose is to retarget any operands that reach it to a new temporal destination.
Where most instructions can reach 32 instructions ahead, the long echo can reach 1024.
For example, \texttt{echo.l =>100} outputs any incoming operands to the 101\textsuperscript{st} upcoming instruction.
The second variant is the \textit{split echo}, which retargets the first two operands it is given to two different destinations.
For example, \texttt{echo =>5, =>10} outputs the first and second incoming operands to the 5th and 10th upcoming instructions, respectively.
This is useful when operands cross basic-block boundaries or after function calls, where all the function's arguments target the first instruction in the function and need to be split out to different consumers.
Any additional operands to the split echo can be either discarded or passed directly to the next instruction (by adding ``\texttt{, =>}'' to the end), which could also be a split echo or a direct consumer.
The previous example discards any operands past the second, while \texttt{echo =>5, =>10, =>} passes them to the following instruction.

\textbf{Duplicate:} As mentioned, the duplication instruction handles the case where an operand is used multiple times.
Any inputs it gets are duplicated, and each copy is retargeted independently.
For example, \texttt{dup =>5, =>10} sends duplicates of each input to the fifth and tenths future instruction.
Since operands that are reused are likely to be reused many times~\cite{franklin1992register}, \texttt{dup} can also create a second duplicate (by adding ``\texttt{, =>}'', resulting in three total outputs) that is passed to the next instruction, which can in turn be another \texttt{dup}.

\textbf{Pick:} The last primary data flow instruction is the \texttt{pick}, which conditionally chooses between two inputs to be output and retargeted.
This is similar to conditional move or set instructions in traditional ISAs.
Forward referencing is complicated if it has to cross basic-block boundaries. 
Additionally, branches and their inevitable branch prediction take up precious hardware resources. 
\texttt{pick} reduces code complexity and the presence of branches.
It also has two variants, one that takes an operand as the condition and one where the condition is an argument (immediate).
This immediate variant is needed to extract one operand from a list of operands and discard the rest.
For example, one loop iteration could target two operands onto an instruction in the following iteration.
When the loop exits, the operands would not target the loop instruction, but some instruction after the loop.
If this is not desired, a pick could retarget one of the desired operands to another instruction.
If only one is needed, the immediate pick would choose it using the hard-coded condition.

\textbf{Nop:} The dedicated \texttt{nop} instruction does nothing except discard any inputs it gets.

\subsection{References in Scry Assembly}

As an example, \autoref{lst-strcpy} shows the C language function \texttt{strcpy} written in Scry assembly.
It starts with a splitting echo instruction that splits the source and destination pointers.
Notice how Scry assembly can also use labels to calculate output references automatically.
Here, \texttt{dup\_dst} refers to the sixth instruction and \texttt{dup\_src} to the second.
Therefore, the first instruction is equivalent to \texttt{echo =>4, =>0}.

The second instruction duplicates the source pointer, sending it to the load instruction as the effective address and to an addition that increments it for the next loop iteration.
Notice the addition on line 9's use of a chain of arrows.
This type of reference mirrors the expected control flow of the program.
\texttt{=>lp\_end=>lp\_start=>dup\_src} signifies that the operand is needed after execution continues to the \texttt{lp\_end} label (the end of the loop), then a branch jumps to \texttt{lp\_start} (the start of the loop), and then is consumed by the instruction at \texttt{dup\_src} (which happens to be the first instruction in the loop.)
This reference is equivalent to \texttt{=>1}, as the instruction at \texttt{lp\_end} is not executed before the jump.
The reference on line 8, \texttt{=>lp\_end=>lp\_start=>dup\_dst} is similarly calculated and is equivalent to \texttt{=>6}.

\begin{lstlisting}[style=scryasmstyle, caption={\texttt{strcpy} function in Scry assembly..}, label=lst-strcpy]
			echo =>dup_dst, =>dup_src
lp_start:
dup_src:	dup =>load, =>inc_src
load:		ld  u8, =>0
			dup =>lp_cond, =>store
lp_cond:	jmp lp_start, lp_end
dup_dst:	dup =>store, =>0
			add.s =>lp_end=>lp_start=>dup_dst
inc_src:	add.s =>lp_end=>lp_start=>dup_src
store:		st
lp_end:		ret return_at
return_at:
\end{lstlisting}

This example also uses the load instruction \texttt{ld}.
It takes two arguments: The data type to be loaded (in this case, an unsigned byte) and an output reference.
It receives one operand from the \texttt{dup} on line 3, which is the absolute address to load from.
It outputs the character loaded to the following \texttt{dup}, which allows it to both be checked against 0 to potentially terminate and be stored at the desired destination by the store instruction, \texttt{st}, on line 10.

The conditional jump instruction, \texttt{jmp}, controls the loop.
It takes two arguments: The label (encoded as an offset) to jump to, if the condition holds, and the label from which to actually jump (called \textit{trigger}) should the condition hold.
The condition is then given as an operand, which here is the loaded character.
Returning from the function is done using the \texttt{ret} instruction, which also takes an argument specifying when the return should trigger.
Since its label (\texttt{return\_at}) immediately follows the instruction, the return will trigger immediately after executing line 11.
The exact semantics of the \texttt{jmp} instruction and function calls and returns are out of the scope of this paper, so will not be further detailed.

\subsection{Output Patterns}

The outputs of Scry instructions come in three different patterns when excluding the above data-flow instructions:

The first is the single output reference we have explained earlier and illustrated in \autoref{lst-flow-example}.
Instructions using this pattern include \texttt{ld} and most of the arithmetic logic unit (ALU) instructions (like \texttt{add.s}).

Some instructions produce no outputs. 
This includes \texttt{st}, \texttt{nop}, \texttt{jmp}, fence (\texttt{fnc}), and trap (\texttt{trp}) instructions.
Other instructions produce one or more outputs, but always pass them to the next instruction.
These instructions are not performance critical or do not take inputs, so they can be reordered easily.
For example, the \texttt{const} instruction produces operands from an encoded field and can always be put exactly before its consumer.
Instructions that do not produce outputs and output only to the following instruction do not need any bits to encode their output reference(s).

Lastly, ALU instructions have their own output format because of their unique requirements.
First, note that while most ALU operations produce one primary output (the result of addition, subtraction, division, etc), some have a secondary output.
For example, addition (e.g., the wrapping ``\texttt{add}''), subtraction, and left-shifts produce carry-bits, multiplication produces high-order bits, and division produces a remainder. 
In the Scry assembly language, the primary output is called \texttt{Low} and the secondary is called \texttt{High}.
For example, addition carry-bits and division remainders are both high outputs.
These high outputs are not always relevant, but often performance critical when needed; the carry or remainder might be checked in an inner loop.
One design decision of Scry is to have as little global state as possible, meaning these high outputs must be treated like any other operands.
Therefore, the ALU instructions that produce two outputs take additional arguments that specify one of the following output variants:
\begin{enumerate}
	\item Both outputs have the same target according to the output reference argument, with the low (e.g., the sum) being first in the order.
	\item Like point 1), but the high (e.g., the carry) is first in the order.
	\item The low follows the output field, while the high is passed to the next instruction.
	\item The high follows the output field, and the low follows the next instruction.
	\item The low follows the output field, while the high is discarded.
	\item The high follows the output field, while the low is discarded.
\end{enumerate}
The output variants are instruction arguments as seen in \autoref{lst-alu-vars} using the same order as the above list.
These six variants of ALU dual-output cover most needs.
If both outputs need independent full references, an \texttt{echo} can be used for fine-grained output retargeting.

\begin{lstlisting}[style=scryasmstyle, caption={The two-output ALU variants.}, label=lst-alu-vars]
	!\gnode{lst-alu-vars-a}!add Low, High, =>10     // Low then High
	!\gnode{lst-alu-vars-b}!sub High, Low, =>10     // High then Low
	!\gnode{lst-alu-vars-c}!mul Low, =>10, High, => // High to next
	!\gnode{lst-alu-vars-d}!nop
	!\gnode{lst-alu-vars-e}!div High, =>10, Low, => // Low to next
	!\gnode{lst-alu-vars-f}!nop
	!\gnode{lst-alu-vars-g}!shl Low,  =>10          // Only Low
	!\gnode{lst-alu-vars-h}!shr High, =>10          // Only High
\end{lstlisting}
\begin{tikzpicture}[remember picture, overlay, green!60!black]
	\path[->, bend right] 
	(lst-alu-vars-a.north west) edge ($(lst-alu-vars-c.west) + (-0.3,0)$) 
	(lst-alu-vars-a.west) edge ($(lst-alu-vars-c.west) + (-0.3,0)$) 
	(lst-alu-vars-b.north west) edge ($(lst-alu-vars-d.west) + (-0.3,0)$) 
	(lst-alu-vars-b.west) edge ($(lst-alu-vars-d.west) + (-0.3,0)$) 
	(lst-alu-vars-c) edge ($(lst-alu-vars-e.west) + (-0.3,0)$) 
	(lst-alu-vars-c.center) edge[bend right=0] (lst-alu-vars-d) 
	(lst-alu-vars-e) edge ($(lst-alu-vars-g.west) + (-0.3,0)$) 
	(lst-alu-vars-e.center) edge[bend right=0] (lst-alu-vars-f) 
	(lst-alu-vars-g) edge ($(lst-alu-vars-h.south west) + (-0.3,-0.1)$) 
	(lst-alu-vars-h) edge ($(lst-alu-vars-h.south west) + (-0.1,-0.1)$) 
	;
\end{tikzpicture}

\subsection{Operand Count Polymorphism}

Polymorphism allows instructions to have different semantics under differing conditions.
Scry heavily uses polymorphism to add semantics to instructions without using any additional encoding space.

Forward-temporal referencing allows instructions to be polymorphic on the number of operands they receive at runtime.
Input count polymorphism changes the semantics of an instruction based on the number of inputs given.
A simple example is the addition instruction, and can also be seen from \autoref{lst-flow-example}:
The first two adds only get one operand each.
However, the third \texttt{add.s} gets two operands.
Using operand count polymorphism, the first two adds become increment operations while the last add is a traditional addition of the two input values.

All Scry ALU instructions may take one or two inputs.\footnote{Support for additional inputs could be added in the future, if deemed beneficial.}
If two inputs are given, the instructions behave normally.
If only one input is given, the second operand is implicit and called an \textit{implicit immediate}.
The value of the implicit operand is chosen to be the most common value that is traditionally used or the most useful one.
For addition and subtraction, ``0'' and ``1'' are the most commonly used immediates~\cite{balakrishnan2003exploiting, ritter1979design, waterman2016design}.
Since adding with zero is equivalent to just using \texttt{echo}, ``1'' was chosen.
Other examples of implicit immediates are ``0'' for the comparison instructions and ``-1'' (all bits set) for \texttt{or} and \texttt{xor} (for setting all bits and flipping all bits, respectively).
More complicated implicit immediates are also possible.
For example, shift instructions naturally take ``1'' as their implicit immediates.
However, multiplication and division most intuitively should take ``2'' as their implicit immediate, performing the same function as the shifts.
Therefore, the implicit immediates for the latter instructions should be different.
The option we use in this paper is to have the pointer size of the platform be the implicit immediate.
This can help reduce the number of instructions needed when doing pointer arithmetic.

Implicit operands are the only available immediate values for ALU instructions.
Having immediate fields in instructions would take up too much encoding space.
This means that any other immediate values that are needed must created using dedicated instructions.
The \texttt{const}/\texttt{grow} instruction pair can create any immediate integer value using twice the number of bits as the value.
I.e., any 8-bit integer can be created using one instruction, 16-bit values need two instructions, etc.

Load and store instructions exploit operand count polymorphism for additional addressing modes.
The first address operand (first operand for loads, second operand for stores---the first being the stored value) is the base address.
If a second address operand is given it functions as a displacement from the base.
This allows for compound pointer arithmetic without additional instructions.
For example, for accessing struct members or array elements.
The exact semantics of the memory instructions will be detailed later.

\subsection{Hardware Considerations}

Forward-temporal referencing puts most of the responsibility for managing operands on the processors themselves.
One challenge for the hardware is managing in-flight operands in long functions and across function calls.
If we do not set limits, the number of operands an instruction might get would be unknown.
It would be difficult to implement a processor that can handle any number of inputs to an instructions.
Scry sets the limit on the number of operands an instruction may consume to four.
Any additional operands that are passed to it are implicitly dropped.
Functions can likewise only take four inputs, with any additional inputs having to be passed on the stack.

The total number of in-flight operands must also be handled with care.
On a function call, the caller's in-flight operands must be managed until the call returns at which point their original targets must still be reached.
The callee function might also have many in-flight operands when it performs another call.
The call stack might therefore have an unknown number of total in-flight operands needing to be managed.
Scry does not limit the total number of in-flight operands, nor the number of in-flight operands in a single function.
Instead, the processor is assumed to be able to store long-lived operands in specially-designed operand caches; which are in turn backed by main memory.
This can be done using a combination of a priority queue to managed short-lived operands~\cite{kohutka2018novel} and dedicated caches that also store the remaining distances an operand has to be stored in its function scope.

\section{Internally Tagged Architecture}
\label{sec-tagged}

The second fundamental Scry design choice is the use of internal tagging~\cite{maroun2025internally}.
A Scry processor associates a type tag with every operand.
The tag specifies the basic type of the operand: Unsigned or signed integer of 1, 2, 4, or 8 bytes.
While Scry does not currently support floating-point types or vectors, the tag will also specify them in the future.
Instructions are polymorphic on input types, so the appropriate operation is chosen based on the tags. 
An addition instruction will choose the integer adder when given integers or the floating-point adder if given floating-point operands.
Using internal tagging to enable instruction polymorphism allows Scry to support many instruction semantics with few encodings, as can be seen by needing only a single encoding to represent addition.

The load instruction is the primary method of setting operand type tags.
Its first argument defines the type to be loaded: \texttt{u8}/\texttt{i8} loads an unsigned or signed byte, \texttt{u16}/\texttt{i16} likewise loads two bytes and so on.
There is also a cast instruction, that changes an operand's type without changing its bits.
The load instruction itself is also polymorphic on input types.
If given an unsigned integer of any size as the first operand, it is zero-extended or truncated to fit the pointer size of the processor and then used as an absolute value (base value) to load from.
If given a signed integer of any size, it is sign-extended and used as a relative address to the load's own address.
This allows the same load instruction encoding to support absolute and relative addressing.
An optional second argument adds a displacement based on its type.
If signed, it is directly added to the base address.
If unsigned, it is treated as an index, scaled to the size of the loaded type, and added to the base address.
These two versions allow both structure member loads and indexed loads, respectively.

Tagging allows Scry to only have one store instruction, \texttt{st}.
It takes no arguments defining the type, as traditional store instructions would.
Instead, it is polymorphic on the type of the first operand (the value to store), ensuring that it stores only the necessary number of bytes.
Its also takes a mandatory second operand and optional third operand that define the address similarly to the load instruction.

\subsection{Type Incompatibility}

As mentioned in~\cite{maroun2025internally}, tagged architectures have a unique challenge in out-of-order (OoO) processors because they may introduce data-dependent exceptions.
The solution in~\cite{maroun2025internally} involved designing the instructions such that their output type is not dependent on the type of operands.
However, Scry has the \texttt{pick} instructions, which chooses its output based on a condition.
If the input types are different, the output type depends on the value of the condition.
This makes it impossible, as mentioned in~\cite{maroun2025internally}, to resolve types in the front-end of an OoO processor.
The result would be that almost every instruction must be speculated for whether it causes a trap because of type incompatibility (e.g., adding an integer and a float).

To remove the need for so much speculation in Scry, Not-a-Result (NaR) values are used \emad{cite something}.
Any instruction that produces at least one operand and experiences an error at runtime will output NaRs instead of the normal result.
NaRs are \textit{poisonous}; instructions that receive at least one NaR input will produce NaR outputs, with few exceptions.
A trap is immediately triggered when a NaR is passed to a store or control-flow instruction.
No other instructions trigger traps, except the dedicated software trap instruction \texttt{trp}.
This includes \texttt{ld}, which produces a NaR if given a NaR.
This solves the problem of OoO processors having to speculate on type-compatibility, but also makes the processor simpler in general, likely benefiting performance.
NaRs have no type.
They contain information that can be used to identify the type of error and maybe where it originated.
This information is passed to the exception handler that is called on a store.
The details of this are undefined and, for now, implementation dependent.

Some operations may frequently produce NaRs, which the program must detect and handle.
In Scry's current design, this includes the \texttt{div} instruction (division by zero) and the load instructions (invalid address).
Extensions will also include floating-point instructions, which already use similar functionality in their not-a-number (NaN), though not identical to the standard~\cite{ieee2019float754}.
Therefore, the \texttt{isnar} ALU instruction is available.
Given any number of inputs, if any are a NaR, the return is true; otherwise, false.
This can be passed to a \texttt{jmp} to handle the exceptional case, like division by zero.

\subsection{Width Agnosticism}

Type polymorphism enables the load and branch instructions, which take address inputs, to not directly care about the pointer size of the processor.
In traditional ISAs, the registers are usually of a size that matches the pointer size, so limit address-needing instructions to support that width.
Scry is unique because it supports any pointer size in the same ISA, whether 16-bit, 32-bit, 64-bit, or beyond.
It can do so because type-knowledge allows the processor to correctly convert any integer into a valid address by either truncating or extending it before performing a memory operation.
Therefore, pointer operands can be of a smaller or larger type than the address space.
A Scry binary will work for any native pointer size, assuming it does not overflow and there are no environmental differences.
This is contrasted with RISC\=/V, which technically describes two binary-incompatible ISAs: RV32 and RV64\cite{waterman2019risc}.\footnote{There are ongoing discussions on a third ISA, RV128.}
Because ALU instructions are type-specific, a 32-bit binary can never be passed a 64-bit address.
In Scry, the binary can be generated to work with any pointer size, and so can be given a 32-bit or 64-bit address by the system when needed.

\section{Encoding}
\label{sec-encoding}

\newcommand{\bb}[2]{\bitbox{#1}[bitheight=3.8ex]{#2}}
\newcommand{\instr}[2]{\bitbox[]{7}{\texttt{#1}} #2}
\begin{figure}
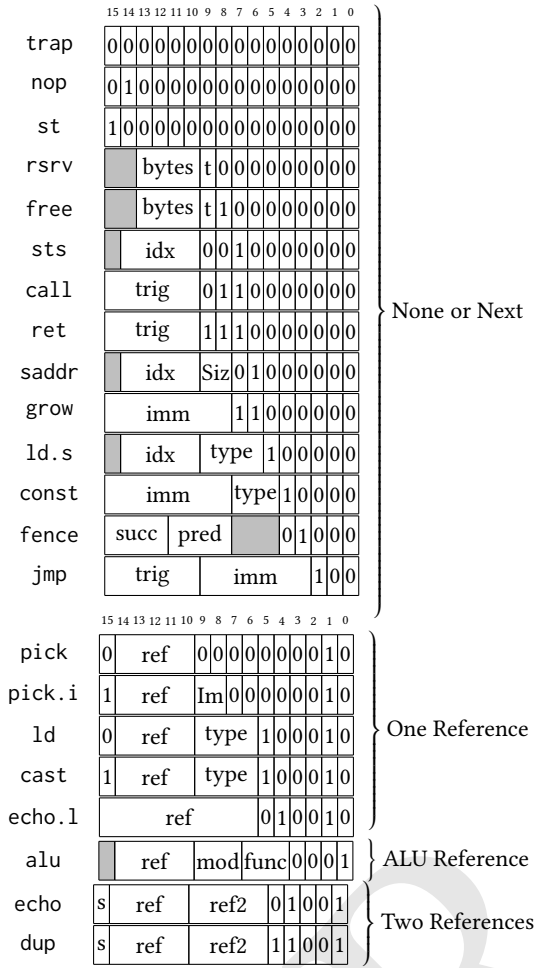

	\vspace{-4ex} 
	\begin{bytefield}[bitheight=4ex]{16}%
		\begin{rightwordgroup}[]{None or Next}%
			\bitbox[]{0}{}\\ 
			$ $ \bitheader{0-15} \\
			\instr{trap}{
				\bb{1}{0} & \bb{1}{0} & \bb{1}{0} & \bb{1}{0} &
				\bb{1}{0} & \bb{1}{0} & \bb{1}{0} & \bb{1}{0} &
				\bb{1}{0} & \bb{1}{0} & \bb{1}{0} & \bb{1}{0} &
				\bb{1}{0} & \bb{1}{0} & \bb{1}{0} & \bb{1}{0}
			}\\
			
			\instr{nop}{
				\bb{1}{0} & \bb{1}{1} &
				\bb{1}{0} & \bb{1}{0} & \bb{1}{0} & \bb{1}{0} &
				\bb{1}{0} & \bb{1}{0} & \bb{1}{0} & \bb{1}{0} &
				\bb{1}{0} & \bb{1}{0} & \bb{1}{0} & \bb{1}{0} &
				\bb{1}{0} & \bb{1}{0}
			}\\
			
			\instr{st}{
				\bb{1}{1} &
				\bb{1}{0} & \bb{1}{0} & \bb{1}{0} & \bb{1}{0} &
				\bb{1}{0} & \bb{1}{0} & \bb{1}{0} & \bb{1}{0} &
				\bb{1}{0} & \bb{1}{0} & \bb{1}{0} & \bb{1}{0} &
				\bb{1}{0} & \bb{1}{0} & \bb{1}{0}
			}\\
			
			\instr{rsrv}{
				\bb{2}{\bitssubclass} & \bb{4}{bytes} & \bb{1}{t} &
				\bb{1}{0} & \bb{1}{0} & \bb{1}{0} & \bb{1}{0} &
				\bb{1}{0} & \bb{1}{0} & \bb{1}{0} & \bb{1}{0} & \bb{1}{0}
			}\\
			
			\instr{free}{
				\bb{2}{\bitssubclass} & \bb{4}{bytes} & \bb{1}{t} & \bb{1}{1} &
				\bb{1}{0} & \bb{1}{0} & \bb{1}{0} & \bb{1}{0} &
				\bb{1}{0} & \bb{1}{0} & \bb{1}{0} & \bb{1}{0}
			}\\
			
			\instr{sts}{
				\bb{1}{\bitssubclass} & \bb{5}{idx} &
				\bb{1}{0} & \bb{1}{0} & \bb{1}{1} &
				\bb{1}{0} & \bb{1}{0} & \bb{1}{0} & \bb{1}{0} &
				\bb{1}{0} & \bb{1}{0} & \bb{1}{0}
			}\\
			
			\instr{call}{
				\bb{6}{trig} & \bb{1}{0} & \bb{1}{1} &
				\bb{1}{1} & \bb{1}{0} & \bb{1}{0} & \bb{1}{0} &
				\bb{1}{0} & \bb{1}{0} & \bb{1}{0} & \bb{1}{0}
			}\\
			
			\instr{ret}{
				\bb{6}{trig} & \bb{1}{1} & \bb{1}{1} &
				\bb{1}{1} & \bb{1}{0} & \bb{1}{0} & \bb{1}{0} &
				\bb{1}{0} & \bb{1}{0} & \bb{1}{0} & \bb{1}{0}
			}\\
			
			\instr{saddr}{
				\bb{1}{\bitssubclass} & \bb{5}{idx} & \bb{2}{Siz} &
				\bb{1}{0} & \bb{1}{1} &
				\bb{1}{0} & \bb{1}{0} & \bb{1}{0} & \bb{1}{0} & \bb{1}{0} & \bb{1}{0}
			}\\
			
			\instr{grow}{
				\bb{8}{imm} & \bb{1}{1} & \bb{1}{1} &
				\bb{1}{0} & \bb{1}{0} & \bb{1}{0} & \bb{1}{0} & \bb{1}{0} & \bb{1}{0}
			}\\
			
			\instr{ld.s}{
				\bb{1}{\bitssubclass} & \bb{5}{idx} & \bb{4}{type} & \bb{1}{1} &
				\bb{1}{0} & \bb{1}{0} & \bb{1}{0} & \bb{1}{0} & \bb{1}{0}
			}\\
			
			\instr{const}{
				\bb{8}{imm} & \bb{3}{type} & \bb{1}{1} &
				\bb{1}{0} & \bb{1}{0} & \bb{1}{0} & \bb{1}{0}
			}\\
			
			\instr{fence}{
				\bb{4}{succ} & \bb{4}{pred} & \bb{3}{\bitssubclass} & \bb{1}{0} & \bb{1}{1} &
				\bb{1}{0} & \bb{1}{0} & \bb{1}{0}
			}\\
			
			\instr{jmp}{
				\bb{6}{trig} & \bb{7}{imm} & \bb{1}{1} &
				\bb{1}{0} & \bb{1}{0}
			}\\
			
		\end{rightwordgroup}
	\end{bytefield}	%
	\begin{bytefield}[bitheight=4ex]{16}%
		\begin{rightwordgroup}{One Reference}
			$ $ \bitheader{0-15} \\
			\instr{pick}{
				\bb{1}{0} & \bb{5}{ref} &
				\bb{1}{0} & \bb{1}{0} & \bb{1}{0} & \bb{1}{0} &
				\bb{1}{0} & \bb{1}{0} & \bb{1}{0} & \bb{1}{0} &
				\bb{1}{1} & \bb{1}{0}
			}\\
			
			\instr{pick.i}{
				\bb{1}{1} & \bb{5}{ref} & \bb{2}{Im} &
				\bb{1}{0} & \bb{1}{0} & \bb{1}{0} & \bb{1}{0} &
				\bb{1}{0} & \bb{1}{0} & \bb{1}{1} & \bb{1}{0}
			}\\
			
			\instr{ld}{
				\bb{1}{0} & \bb{5}{ref} & \bb{4}{type} & \bb{1}{1} &
				\bb{1}{0} & \bb{1}{0} & \bb{1}{0} & \bb{1}{1} & \bb{1}{0}
			}\\
			
			\instr{cast}{
				\bb{1}{1} & \bb{5}{ref} & \bb{4}{type} & \bb{1}{1} &
				\bb{1}{0} & \bb{1}{0} & \bb{1}{0} & \bb{1}{1} & \bb{1}{0}
			}\\
			
			\instr{echo.l}{
				\bb{10}{ref} &
				\bb{1}{0} & \bb{1}{1} & \bb{1}{0} & \bb{1}{0} & \bb{1}{1} & \bb{1}{0}
			}
		\end{rightwordgroup}
	\end{bytefield}	%
	\begin{bytefield}[bitheight=4ex]{16}%
		\begin{rightwordgroup}{ALU Reference}		
			\instr{alu}{
				\bb{1}{\bitssubclass} & \bb{5}{ref} & \bb{3}{mod} & \bb{3}{func} &
				\bb{1}{0} & \bb{1}{0} & \bb{1}{0} & \bb{1}{1}
			}
		\end{rightwordgroup}
	\end{bytefield}	%
	\begin{bytefield}[bitheight=4ex]{16}%
		\begin{rightwordgroup}{Two References}	
			\instr{echo}{
				\bb{1}{s} & \bb{5}{ref} & \bb{5}{ref2} &
				\bb{1}{0} & \bb{1}{1} & \bb{1}{0} & \bb{1}{0} & \bb{1}{1}
			}\\
			
			\instr{dup}{
				\bb{1}{s} & \bb{5}{ref} & \bb{5}{ref2} &
				\bb{1}{1} & \bb{1}{1} & \bb{1}{0} & \bb{1}{0} & \bb{1}{1}
			}
		\end{rightwordgroup}
	\end{bytefield}	%
	\caption{Encoding of Scry instructions, with their mnemonic on the left and output pattern on the right.}
	\label{fig-encoding}
	\vspace{-15pt}
\end{figure}

\addtolength{\tabcolsep}{-0.4em}
\begin{table}
	\caption{ALU instruction encodings. Dashes signify unused encodings.}
	\label{tab-alu}
	\centering
	\begin{tabular}{lllll}
		\toprule
		func & mod & Mnemonic   & Operation & Implicit\\
		\midrule
		000 & 000  & \texttt{eq}     & Equal comparison & 0 \\
		& 111  & \texttt{add.s}    & Addition, saturating & 1 \\
		& mod  & \texttt{add}  & Addition, wrapping, with carry			& 1\\
		\midrule
		001 & 000  & \texttt{and}    & Bitwise AND & 1 \\
		& 111  & \texttt{sub.s}    & Subtraction, saturating & 1\\
		& mod  & \texttt{sub}   & Subtraction, wrapping, & 1\\
		&   &    & with carry & 1\\
		\midrule
		010 & 000  & \texttt{lt}     & Less-than comparison & 0 \\
		& 111  & \texttt{gt}     & Greater-than comparison & 0\\
		& mod  & \texttt{shl}     & Shift left with carry & 1\\
		\midrule
		011 & 000  & \texttt{or}     & Bitwise OR & All bits set \\
		& 111  & \texttt{xor}    & Bitwise XOR  & All bits set  \\
		& mod  & \texttt{shr}     & Shift right with carry & 1\\
		\midrule
		100 & 000  &  \texttt{isnar}  & Whether inputs include NaR & - \\
		& 111  &  -      & - & - \\
		& mod  & \texttt{mul}    & Full multiplication & \texttt{usize}\\
		\midrule
		101 & 000  & -      & - & - \\ 
		& 111  & -      & - & - \\ 
		& mod  & \texttt{div}    & Division with remainder & \texttt{usize} \\
		\midrule
		110 & -  & -      & - & - \\ 
		\midrule
		111 & -  & -      & - & - \\ 
		\bottomrule
	\end{tabular}
	\vspace{-15pt}
\end{table}
\addtolength{\tabcolsep}{0.4em}

Scry encodes all instructions in two bytes in little-endian.
\autoref{fig-encoding} shows the encoding for all instructions.
To maximize efficiency, the encoding is irregular compared to traditional ISAs with no instruction formats.
However, common fields are shared between different instructions.
For example, the \texttt{ref} field is used by single-output and ALU-instructions and is encoded using bits 10-14.
The \texttt{type} field is used by instructions that set or change a value's type, like \texttt{ld} or \texttt{cast}.

The instructions are grouped into four output types.
The instructions that either produce no outputs or produce outputs directly to the next instruction use ``00'' in their two least significant bits.
Next are the instructions that produce one output with a regular output reference (``10'' in the least significant bits).
This includes the long echo, which is only different because its reference is ten bits.
The ALU instructions (``0001'' in their least significant bits) also have a reference field, but depending on the other fields, they might also produce outputs to the next instruction.
Lastly, the splitting echo and duplicate (``1001'' in their least significant bits) each have two reference fields and one \texttt{s} flag, signifying whether they also output to the next instruction.
Encodings with ``11'' in their two least significant bits are currently unused.

Encoding output patterns in groups enables the tracking of operand counts independent of executing individual instructions.
This is necessary because many instructions are operand-count polymorphic, and their exact variant must be determined early.
This encoding scheme ensures that each instruction has a statically known output count for each input count.
As such, input-operand counts can be tracked in early stages of a pipeline to ensure that operand-count polymorphism can be resolved early.
This would be especially necessary for OoO implementations.
The \texttt{call} instruction initially seems to counteract this design.
Any operands a function returns implicitly target the instruction following the call instruction.
The number of operands returned depends on the function's body, so it cannot be statically determined.
However, processors see the instruction sequence in the callee as directly preceding the code following the call.
As such, its operand-count tracking in the callee is also valid following the return, meaning \texttt{call}/\texttt{ret} do not pose a problem regarding operand-count tracking.

ALU instructions use a unique encoding scheme to minimize encoding space use.
Having two output fields wastes encoding space when either the high or low outputs are not needed.
Instead, ALU instructions have a 5-bit primary output field, a 4-bit function field (\texttt{func}), and a 3-bit modifier field (\texttt{mod}).
Any instruction where \texttt{mod}'s bits are all clear or all set can only be an operation that produces one output using an output reference.
Examples include the saturating addition (where no carry is produced, seen in \autoref{lst-flow-example}) or saturating subtraction instructions, the logical comparison instructions, or the bitwise logical instructions.
For the remaining cases of \texttt{mod}, the field behaves as an output modifier for an operation; e.g., a wrapping addition with a carry.

This ALU encoding scheme provides all necessary features but with minimal encoding cost.
The combination of the \texttt{func} and \texttt{mod} fields enables encoding 24 ALU variants: 16 single-output variants and 8 two-output variants.
A minimal ISA implementation will need only 14 of those variants to encode the most important ALU instructions:
2 additions and subtractions (wrapping with carry and saturating without), multiplication and division instructions (both two-output), six single-output logical instructions (\texttt{and}, \texttt{or}, \texttt{xor}, \texttt{lt}, \texttt{gt}, \texttt{eq}), and shifting left and right (both with carries).

\autoref{tab-alu} shows the assignments of ALU instructions to the \texttt{func} and \texttt{mod} fields.
When $\texttt{mod}=0$ or $\texttt{mod}=111$, the instruction has a single output; e.g., the saturating addition ``\texttt{add.s}''.
Otherwise, the instruction has two outputs using the previously mentioned output variants; e.g., the wrapping addition ``\texttt{add}''.
The implicit operands of the \texttt{mul} and \texttt{div} instructions are the pointer size of the platform and so are implementation-defined.
For example, a platform with a 32-bit memory space would use ``4'' for these implicit operands.
All the implicit operands' type is identical to the type of the first operand.
\texttt{isnar} does not use an implicit operand, as it can check for NaR within any input operands, even if given only one.

For brevity, we have omitted explaining all intructions in this paper in detail.
Instead, we give a quick overview of those that have yet to be discussed: 

\begin{itemize}
	\item \texttt{rsrv}/\texttt{free}: Scry has first-class support for the program stack. These two instructions allow a function to reserve or free additional space on their stack frames.
	\item \texttt{ld.s}/\texttt{st.s}: These instructions are analogous to the regular load and store instruction, but directly target the stack frame instead.
	\item \texttt{saddr}: Scry has no accessible register for tracking the stack pointer. Instead, this instruction can be used to get the address of any data on the stack frame.
	\item \texttt{fence}: A memory fence instruction similar to RISC-V's \texttt{fence} instruction \cite{waterman2019risc}. Ensures memory access ordering is maintained by the processor when needed.
\end{itemize}

\section{Evaluation}
\label{sec-evaluation}

We have implemented an ISA simulator and an assembler supporting Scry.
However, because of Scry's significant fundamental difference from traditional ISAs, implementing a compiler and a processor is a significant challenge we have yet to undertake.
Therefore, we perform two types of evaluation:
Compare the encoding efficiency of the Scry ISA against that of RISC\=/V and compare machine-code implementations of Scry functions against those of RISC\=/V.

\subsection{Encoding Efficiency}

We calculate an instruction's encoding space usage by looking at how many \textit{code points} it uses.
An instruction like \texttt{nop} uses only one code point, as it has no fields.
However, the \texttt{lt} instruction has two fields of five and four bits for the output and type arguments, respectively. 
Therefore, it uses $2^{5+4} = $ \numprint{\fpeval{2^9}} code points.
RISC\=/V's compressed instructions effectively have 16 bits they cannot use, meaning we add them as a field to the calculation.
For example, the \texttt{c.ebreak} has no fields and \texttt{c.add} has two register fields~\cite{waterman2019risc}.
Therefore, they occupy $2^{16}=$ \numprint{\fpeval{2^16}} and $2^{5+5+16}=$ \numprint{\fpeval{2^26}} code points, respectively.

We compare the Scry encoding efficiency to that of RISC\=/V by summing the amount of encoding space they each use.
Cumulatively, Scry uses \numprint{\ScryPoints} code points.
This is \scryOccupancy \% of the available 16-bit encoding space.
In contrast, RV64IMC uses \numprint{\riscvPointsRelevant} code points, which is \riscvOccupancy \% of the 32-bit encoding space.\footnote{The RISC\=/V analysis is based on the instruction encoding data given in: \\ \href{https://github.com/riscv/riscv-opcodes/commit/b30cec9}{https://github.com/riscv/riscv-opcodes/commit/b30cec9}}
As Scry will need fewer instruction additions when adding extensions like floating point support (because of internal tagging and polymorphism), the Scry encoding is orders of magnitude more efficient than RISC\=/V and much more extensible.

\subsection{Assembly Programs}

We evaluate the proposed Scry ISA using only hand-written assembly programs.
We compare their structure to the same programs written in C and compiled using the GCC RISC\=/V compiler.
Writing assembly programs in Scry is more challenging than traditional ISA because of the difficulty of managing references.
Therefore, we limit the programs to small programs, with none above 40 Scry instructions.
We choose programs from the C standard library that are potentially simplified with open-source C implementations to represent real-world use cases.

We have implemented four functions from the C standard library that are small and likely to be often used.
\texttt{strcpy} and \texttt{memcpy} copy either a null-terminated string or a simple memory block to a target destination.
Their Scry implementation is shown in \autoref{lst-strcpy} and \autoref{lst-memcpy}.
These core functions are often used and must have good performance.
\texttt{isxdigit} (\autoref{lst-isxdigit}) checks whether a given character represents a hexadecimal digit.
While not as often used, it is branch-heavy in traditional ISAs because of the checks against the ASCII character table.
Intuitively, Scry struggles more than traditional ISAs around branches because of forward-temporal referencing.
The last is \texttt{bsearch}, which uses the binary search algorithm to find a value in an array.
This is the largest program we have implemented in Scry and should give some insight into the overall performance of the ISA.

We have also implemented three non-standard functions.
First, \texttt{cmpu8} implements a comparison of unsigned bytes for use with \texttt{bsearch}.
This complements the evaluation of \texttt{bsearch}, which always takes a comparison function.
We have chosen this function because of its simplicity and because it is likely one of the most used comparisons with \texttt{bsearch}.
The \texttt{find\_max} function does a simple iterative search through an array, returning the largest value.
This exhibits loop iteration based on a counter, a common pattern not covered by the other functions.
Lastly, \texttt{hextol} simplifies the C standard library \texttt{strtol} function.
The latter parses a string of any base into a long integer.
\texttt{hextol} only parses hexadecimal numbers, does not handle leading whitespaces, and does not return the end pointer.
This exhibits loops and consecutive if/else statements.

\begin{lstlisting}[style=scryasmstyle, caption={\texttt{memcpy} function in Scry assembly.}, label=lst-memcpy]
			echo =>dup_source, =>dup_sink, =>
			dup =>check_zero, =>dec_count
check_zero:	jmp lp_end, 0	
dup_source:	dup =>load_next, =>inc_source

lp_start:
dec_count:	sub Low, =>0
			dup =>lp_cond, 
				=>lp_end=>lp_start=>dec_count
load_next:	ld u8, =>store_copy
lp_cond:	jmp lp_start, lp_end
dup_sink:	dup =>store_copy, =>inc_sink
inc_source:	add Low, =>0
			dup =>lp_end=>lp_start=>load_next,
				=>lp_end=>lp_start=>inc_source
inc_sink:	add Low, =>lp_end=>lp_start=>dup_sink
store_copy:	st
lp_end:		ret 0
\end{lstlisting}

\begin{lstlisting}[style=scryasmstyle, caption={\texttt{isxdigit} function in Scry assembly.}, label=lst-isxdigit]
				dup =>sub_0, =>without_bit5
				ret return
				const u8, 48
sub_0:			sub Low, =>lt_10
				const u8, 10
lt_10:			lt =>dig_or_let
				const u8, 223
without_bit5:	and =>sub_a
				const u8, 65
sub_a:			sub Low, =>lt_6
				const u8, 6
lt_6:			lt =>dig_or_let
dig_or_let:		or =>0
return:
\end{lstlisting}

We compare the static composition of the selected functions.
\autoref{tab-results} shows, for each function, the total number of instructions, the number of instruction bytes, the number of data-flow instructions, the number of control-flow instructions, and the number of instructions comprising the function prologue or epilogue.
Data-flow instructions' only purpose is data movement within the processor.
For Scry, this includes \texttt{echo}, \texttt{dup}, \texttt{pick}, and \texttt{const}.

For the smallest functions, we see that Scry and RISC\=/V require comparable byte numbers, but Scry requires more instructions.
\texttt{strcpy} and \texttt{cmpu8} differ only by 2 bytes.
\texttt{memcpy} needs six more instructions in the Scry version than RISC\=/V; however, \texttt{find\_max} needs eight more in the RISC\=/V version than Scry.
Scry's static encoding density shows promise for the rest of the functions, needing 14, 24, and 40 fewer bytes for \texttt{isxdigit}, \texttt{hextol}, and \texttt{find\_max}, respectively.
In \texttt{bsearch}'s case, the benefit comes from the prologue and epilogues, where Scry only needs seven instructions while the RISC\=/V implementation needs 25.
The savings come from Scry not needing to save and reload registers and manage the stack pointer (a single instruction is used to free the stack).
Looking at the amount of data-flow instructions, we see that Scry uses many more than RISC\=/V, as expected.
However, this does not come at the cost of the instruction density.

The trend these numbers show is that small Scry functions are comparable in density to RISC\=/V, but the larger and more complicated the function becomes, the more likely Scry is to be denser.
Note that these functions are so simple that no register spilling is needed in the RISC\=/V versions.
Larger functions with many operands will affect RISC\=/V's density adversely.
Scry will never need to add instructions for spilling, so it will not be affected as much.\footnote{At runtime, a Scry processor might still need to spill long-lived operands to cache/memory in the background, which will affect execution times by using cache/memory bandwidth.}
Lastly, Scry's use of \texttt{pick}s instead of branches for small-scale decision-making successfully reduced the number of control-flow instructions without adversely affecting encoding efficiency.

Five of these functions comprise a loop doing the main body of the work.
\autoref{tab-results-main-loop} shows statistics on the number of instructions in the main loop only, the number of bytes, the number of conditional branches.
Almost all functions use slightly more instructions and bytes in their main loop in the Scry version than RISC\=/V.
Scry's density, therefore, does not extend to the most heavily executed code.
However, loops as small as those analyzed certainly fit in any first-level cache, so the byte density is of little concern.
The total number of bytes in a function is of higher concern, since that will increase the cache miss rate.
Many of these functions are called repeatedly and often, so their reduced size in Scry will reduce cache pressure.
Looking at the conditional branches, we see that Scry successfully minimizes the need for branches.
Only a single case needs more than one conditional branch to manage the loop.
The biggest benefit is in \texttt{hextol}, where RISC\=/V needs four branches (to check for digits, capital/small letters, and zero) but Scry needs only one.
Scry uses \texttt{jmp} to check for the terminator, while extracting the character value only uses \texttt{pick}s.
This increases the loop size, but only marginally.
The benefit to branch prediction efficiency will likely supersede the four additional bytes.

\newlength{\instrlen}
\settowidth{\instrlen}{Instructions}
\setlength{\instrlen}{\dimexpr(\instrlen-2\tabcolsep)/2}

\newlength{\cntllen}
\settowidth{\cntllen}{Control}
\setlength{\cntllen}{\dimexpr(\cntllen-2\tabcolsep)/2}

\newlength{\brlen}
\settowidth{\brlen}{Branches}
\setlength{\brlen}{\dimexpr(\brlen-2\tabcolsep)/2}

\begin{table}
	\caption{Instruction composition for evaluated programs. Left values are Scry, right are RISC\=/V.}
	\label{tab-results}
	\centering
	\begin{tabular}{l | >{\centering\arraybackslash}p{\instrlen} : >{\centering\arraybackslash}p{\instrlen} | c : c | c : c | >{\centering\arraybackslash}p{\cntllen}  : >{\centering\arraybackslash}p{\cntllen} | c : c }
		\toprule
		Function & \multicolumn{2}{c}{Instructions} & \multicolumn{2}{c}{Bytes} & \multicolumn{2}{c}{Data} & \multicolumn{2}{c}{Control} & \multicolumn{2}{c}{Logues} \\
		\midrule
		\texttt{strcpy} & 10& \textbf{7} & 20&\textbf{18} & 4&\textbf{1} & 2&2 & 2&\textbf{1} \\
		\texttt{memcpy} & 14& \textbf{9} & 28&\textbf{22} & 6&\textbf{1} & 3&3 & 3&\textbf{1} \\
		\texttt{isxdigit} & 13&13 & \textbf{26}&40 & 6&\textbf{4} & \textbf{1}&3 & \textbf{2}&3 \\
		\texttt{bsearch} & \textbf{33}&47 & \textbf{66}&106 & 15&\textbf{9} & \textbf{5}&7 & \textbf{7}&25\\
		\texttt{cmpu8} & 5& \textbf{4} & \textbf{10}&12 & 1&\textbf{0} & 1&1 & 2&\textbf{1} \\
		\texttt{find\_max} & 14&\textbf{13} & \textbf{28}&36 & 9&\textbf{5} & \textbf{2}&5 & 2&2 \\
		\texttt{hextol} & 36&\textbf{31} & \textbf{72}&96 & 21&\textbf{8} & \textbf{2}&6 & 2&\textbf{1} \\
		\bottomrule
	\end{tabular}
	\vspace{-10pt}
\end{table}

\begin{table}
	\caption{Main loop composition for evaluated programs. Left values are Scry, right are RISC\=/V.}
	\label{tab-results-main-loop}
	\centering
	\begin{tabular}{l | >{\centering\arraybackslash}p{\instrlen} : >{\centering\arraybackslash}p{\instrlen} | c : c | >{\centering\arraybackslash}p{\brlen} : >{\centering\arraybackslash}p{\brlen}}
		\toprule
		Function & \multicolumn{2}{c}{Instructions} & \multicolumn{2}{c}{Bytes} & \multicolumn{2}{c}{Branches} \\
		\midrule
		\texttt{strcpy} & 8&\textbf{5} & 16&\textbf{14} & 1&1 \\
		\texttt{memcpy} & 9&\textbf{6} & 18&\textbf{16} & 1&1 \\
		\texttt{bsearch} & 20&\textbf{12} & 40&\textbf{34} & \textbf{2}&3  \\
		\texttt{find\_max} & 10&\textbf{7} & \textbf{20}&22 & \textbf{1}&2  \\
		\texttt{hextol} & 32&\textbf{18} & 64&\textbf{60} & \textbf{1}&4 \\ 
		\bottomrule
	\end{tabular}
	\vspace{-10pt}
\end{table}

\section{Conclusion}
\label{sec-conclusion}

We have presented the Scry ISA and two of its primary design elements
Forward-temporal referencing specifies when instruction outputs are consumed, with instruction inputs being implicit.
Internal tagging has the processor track data types instead of instruction.
Operand count polymorphism and type polymorphism enable rich semantics that usually require additional instructions.
We presented a proof-of-concept encoding of the Scry ISA using instructions with the same feature set as RISC\=/V's RV64IMC.
Our results showed orders of magnitude better encoding efficiency, with Scry needing only \numprint{\ScryPoints} code points compared to RISC\=/V's \numprint{\riscvPointsRelevant}.
Static instruction density was evaluated using hand-written Scry programs, showing that small functions have comparable instruction density but that larger programs use fewer instruction bytes in Scry.

These numbers do not tell the whole story without an implemented compiler and processor.
Therefore, further work must implement a compiler that can target the Scry ISA, and a processor must be built to run the programs to make real-world comparisons.


\printbibliography

\end{document}